\documentclass[namedreferences]{kluwer}
\usepackage{psfig}

\newcommand{\citeN}[1]{\citeauthor{#1}\ (\citeyear{#1})}
\newcommand{\citeNP}[1]{\citeauthor{#1},\ \citeyear{#1}}

\newcommand{\apj}{{\it Astrophys.~J.}}

\newcommand{\aap}{{\it Astron.~Astrophys.}}

\newcommand{\mnras}{{\it Monthly Notices Roy.\ Astron.\ Soc.}}

\newcommand{\solphys}{{\it Solar~Phys.}}

\newcommand{\ujlisti}{
\itemsep=0 em
\parsep=0.5 em
\partopsep=0.25 em
\topsep=0 em}
\newcommand{\ujlistii}{
\itemsep=0 em
\parsep=0.5 em
\partopsep=0.25 em
\topsep=0 cm}

\newenvironment{lista}{\begin{list}{--}{\ujlisti}}{\end{list}}

\newcommand{\vc}[1]{\mbox{\bf #1}}

\newcommand{\Reyno}{\mbox{Re\/}}
\newcommand{\Strouno}{\mbox{St\/}}

\begin{document}                                                                                   
\begin{article}
\begin{opening}

\runningtitle{Turbulence in the Solar Photosphere}
\runningauthor{Petrovay}

\title{TURBULENCE IN THE SOLAR PHOTOSPHERE}

\author{K. \surname{Petrovay}}
\institute{E\"otv\"os University, Dept.~of Astronomy, Budapest, Pf.~32,
	   H-1518 Hungary, and\\
	   Instituto de {Astrof\'\i sica} de Canarias, 
	   La Laguna, Tenerife, E-38200 Spain
	   } 

\date{[{\it Space Science Reviews}, {\bf  95}, 9--24 (2001)]}

\begin{abstract} 
The precise nature of photospheric flows, and of the transport effects they give rise
to, has been the subject of intense debate in the last decade. Here we attempt
to give a brief review of the subject emphasizing interdisciplinary (solar
physics---turbulence theory) aspects, key open questions, and recent
developments.
\end{abstract}
\end{opening}

  



\section{Introduction}

The stochastic nature of the motions observed on the Sun and their enormous
Reynolds numbers suggest that the photospheric plasma must be in a strongly
turbulent state. On the other hand, the extreme physical conditions prevailing
in the photosphere (strong inhomogeneity and anisotropy, non-diffusive
radiative transfer, penetrative convective motions in a stably stratified
layer, compressibility effects, intermittency, etc.) imply that photospheric
motions differ fundamentally from the textbook case of isotropic incompressible
Kolmogorov turbulence. The precise nature of these flows, and of the transport
effects they give rise to, has been the subject of intense debate in the last
decade. The contradictory views formulated on the topic range from na\"{\i}ve
and uncritical applications of the Kolmogorov theory to the outright denial of
the turbulent nature of photospheric flows. 

The aim of this paper is not to give an exhaustive review or to cite all 
relevant papers. Instead, the emphasis will be on  interdisciplinary aspects,
key open questions, and recent developments.

\section{The nature of photospheric flows}

\subsection{MHD turbulence: a summary of theory and experiments}

\subsubsection{Phenomenology: shear instability, cascade, inertial range}

In a very simplistic description that still grasps the essence, turbulence may
be regarded as the result of shear flow instability. Shear instabilities are 
present in a very wide class of flows. In the nonlinear regime the unstable
modes develop into vortices with characteristic scales similar to the length
scale $L$ of the shear of the original velocity field. The instability develops
on a timescale $L/V$, $V$ being the scale of the original velocity difference
between the shearing layers. The instability condition is that the
Reynolds number, $\Reyno=LV/\nu$ with $\nu$ the viscosity, exceed some critical
value.

For a sufficiently high Reynolds number the vortices originating in the
instability are on their turn also susceptible to shearing instabilities,
whereby a whole hierarchy of vortices, each level characterized by a length
scale $l$ and some characteristic velocity $v$, may develop for high Reynolds
numbers. The size of the smallest eddies is set by the condition that their
effective Reynolds number $lv/\nu$ should be about equal to the critical value
so they can remain stable. Vortices at each level draw their energy from
vortices of the next higher level by means of the instability; hence, the
mechanical energy input at the largest scale $L$ may be thought of 
``cascading'' down to smaller and smaller scale motions until at the shortest
scales it is dissipated by viscosity. 

It must be emphasized that the above scenario is simplistic almost to the point
of being erroneous: the series of instabilities leading to turbulence is far
more complex than outlined above, and details vary greatly among flow types
(\citeNP{Drazin+Reid}). Yet this classic phenomenological picture of the origin
of turbulence may be considered valid as a common denominator of the wide
variety of actual processes taking place in various flows. Indeed, at this
basic level it is independent of the mechanism generating the motion at the
largest scales, i.e. for smaller scale turbulence it should be equally valid
for turbulent convection or for random driving. At the same time one must be
aware that this picture overlooks some important features of turbulence such as
intermittency and coherent structures (see below).

In a more mathematical formulation, the hierarchy of ``eddies'' or ``vortices''
can be represented by a Fourier decomposition of the flow field. The wavenumber
$k\sim 1/l$ then takes the place $l$ as independent variable. The total kinetic
energy of the flow may be regarded as the sum or integral of contributions from
each level of the hierarchy, $E=\int E_k\,dk$. 

Implicit in the above cascading scenario is the idea that the energy transfer is
{\it local\/} in wavenumber space: each vortex draws its energy from vortices
only one step higher up in the hierarchy. If the (large) energy input scale and
the (small) viscous scale are separated by many orders of magnitude, one may
thus expect that the local energy transfer between modes is ``ignorant'' of the
driving processes at large scales and of the viscous processes at small scales
and its dynamics is only determined by the inertial term ($\vc v\nabla)\vc v$ of
the equation of motion. Hence, the shape of the spectrum in this {\it inertial
range\/} cannot depend on any imposed length scale: the spectrum $E_k$ is 
expected to be scale-independent, i.e.\ a power law. ``One step'' in
the hierarchy then translates to one decade (or any similar fixed factor) in
wavenumbers, as the vortex hierarchy is self-similar. The kinetic energy of
eddies at one level is thus $v^2/2=\int_{k/10}^k E_k\,dk\sim kE_k$.

\subsubsection{Kolmogorov vs.\ Iroshnikov--Kraichnan spectra}

In a stationary state where the energy input by the driving balances
dissipation, the spectral energy transfer rate $\epsilon$ from low to high
wavenumbers must be independent of $k$ (and equal to the dissipation rate) in
the inertial range where no significant input or dissipation occurs: 
\begin{equation} 
\epsilon\equiv v^2/2\tau =\,\mbox{const.}          \label{eq:transfer}          
\end{equation} 
where $\tau$ is the transfer timescale. In the scenario outlined above the
transfer is due to the shear instabilities developing on larger eddies so its
timescale is related to the growth rate of those instabilities: $\tau\sim
l/v\sim(vk)^{-1}$. Substituting this into equation (\ref{eq:transfer}) we find
$v\sim\epsilon^{1/3}k^{-1/3}$ or
\begin{equation} 
E_k\sim \epsilon^{2/3}k^{-5/3} ,    \label{eq:Kolmo}  
\end{equation} 
the famous Kolmogorov spectrum. There is ample evidence from experiments and
observations for a Kolmogorov range in neutral fluids.

\citeN{Iroshnikov} and \citeN{Kraichnan:IK} noted that the relation $\tau\sim
l/v$ will not necessarily hold in a {\it conducting\/} fluid. In the presence
of a magnetic field with an associated Alfv\'en speed exceeding $v$, the
turbulent fluctuations should behave as Alfv\'en wave packets travelling along
field lines with the Alfv\'en speed (the so-called Alfv\'enic effect). This
should be the relevant case even in the absence of a large-scale mean field as
the amplitude of turbulent magnetic fluctuations (produced by a small-scale
dynamo, cf.\ Sect.~3.1) on the largest scales is high enough to dominate
smaller-scale motions for all plausible spectral exponents. Iroshnikov and
Kraichnan further argued that the interaction time of two wave packets of size
$l$, travelling in the opposite direction, is  $\tau_i=l/v_A\ll l/v$. In such a
short time only a small amount of  energy, say $\delta E$ can be fed
from one mode into the other. The number of such random interactions needed to
change the energy of a wave packet significantly is $(E/\delta E)^2=(v_A/v)^2$. 
Multiplying $\tau_i$ with this number one finds $\tau\sim(E/\delta E)^2\tau_i =
(v_A/v) (l/v)$, where the Alfv\'en speed $v_A$ is determined by the large-scale
magnetic field, and is thus  independent of $l$. When substituted into
(\ref{eq:transfer}), this results in $v\sim(\epsilon v_A)^{1/4} k^{-1/4}$ or 
\begin{equation}
 E_k\sim (\epsilon v_A)^{1/2} k^{-3/2}  ,  \label {eq:IK}   
\end{equation}
known as the Iroshnikov--Kraichnan (IK) spectrum. Note that  
$\epsilon=\,$const.\ implies a kinetic energy transfer rate independent of $k$
while in the MHD case it is not the kinetic but the total (kinetic$+$magnetic)
energy that is conserved. Similar arguments (\citeNP{Biskamp}) can be used to
show that if there is no cross-helicity the kinetic and magnetic energy spectra
will be in equipartition in the inertial range, so their transfer rates are
equal and constant. 

In recent years doubts have arised regarding the validity of the IK scaling in
MHD turbulence. On the theoretical side, \citeN{Goldreich+Sridhar} pointed out
some flaws in the simplified treatment of wave interactions in the IK analysis,
proposing an alternative model. \citeN{Politano+:struct.fn} derived exact
scaling relations for certain third-order structure functions: the exponents of
these scalings differ from those predicted by the IK model for similar (though
not identical) correlations, suggesting (though not proving) that the IK
scaling may not be correct. These findings are supported by the numerical
simulations of \citeN{Muller+Biskamp:IK} where instead of an IK scaling
a standard Kolmogorov range was found in decaying 3D MHD turbulence with
vanishing cross-helicity. The fact that the numerical values of the exponent
(1.5 viz.\ 1.66) are quite close for the two scalings makes it difficult to
distinguish between the two cases on the basis of empirical data (observations
or experiments). We may thus conclude that the jury is still out on the problem
of the shape of the inertial range energy spectrum in MHD turbulence. In fact,
owing to departures from locality of the transfer in wave number space, the
energy spectrum may not even be power law throughout (\citeNP{Biskamp}).

\subsubsection{Inverse cascades}

As we have seen, the overall tendency is for the energy to cascade towards
small scales. This is generally also true for other ideal invariants.
However, if the spectra of two conserved quantities are not completely
independent their simultaneous direct cascade may be inhibited so that one of
them cascades towards large scales: this is known as an {\it inverse cascade.}

Let the spectrum $H_k$ of a conserved quantity $H$ be related to $E_k$ as
$H_k=h(k) k^n E_k$ where $h(k)$ has an upper or lower bound (or both). As a
concrete example we may take the magnetic helicity $H=\vc A\vc B$ with $\vc A$
the vector potential and $\vc B$ the magnetic field strength, conserved for 3D 
ideal MHD turbulence; here, $n=-1$, and the relative helicity
$h(k)$ is bounded from above.  Taking into account equation
(\ref{eq:transfer}), the transfer rate of $H$ normalized by $\epsilon$ is 
\begin{equation}
 kH_k/\epsilon\tau =h(k)k^n \label{eq:inverse}   
\end{equation}
The constancy of this transfer rate constitutes no problem if $n<0$ and $h(k)$
has a  lower bound or $n>0$ and $h(k)$ has an upper bound (this latter being the
case for the kinetic helicity $\vc u\nabla\times\vc u$, an invariant in 3D
ideal hydrodynamic turbulence). But in the case $n<0$ with $h(k)$ bounded from
above,  such as in the case of magnetic helicity, the transfer rate
(\ref{eq:inverse}) can clearly not remain constant: in the high Reynolds number
limit only a vanishingly small fraction of $H$ fed in at the input scale can be
dissipated at the small scales. Most of the input of $H$ has then no choice but
to cascade towards {\it low\/} $k$ values. Therefore, magnetic helicity suffers
an inverse cascade in 3D MHD turbulence. In fact, this inverse cascade of the
magnetic helicity is responsible for the large-scale dynamo effect.

Finally, in the case if $n>0$ and $h(k)$ is bounded from below equation
(\ref{eq:inverse}) would predict a transfer rate {\it increasing\/} with $k$
which, in a stationary state, is clearly an impossibility in the inertial range
where no driving is present. In this situation the only way to preserve the
conservation of both the energy and $H$ is to assume $\epsilon\neq\,$const.\ in
the inertial range; instead, $\epsilon$ should decrease with $k$ so $H$ may show
a direct cascade with a constant transfer rate there. In analogy with the
previous case it is now the {\it energy\/} that must show an inverse cascade.
This case is realized in 2D hydrodynamic turbulence where the enstrophy
$(\nabla\times\vc u)^2/2$ is an ideal invariant and its spectrum is simply
$k^2E_k$ i.e. $n=2$ and $h(k)\equiv 1$ in this particular case. 

\subsubsection{Real-world turbulence vs.\ textbook turbulence}

\begin{quote}

\footnotesize 
\strut\\
``Turbulence is a dangerous topic which is often at the origin of serious fights
in the scientific meetings devoted to it since it represents extremely different
points of view, all of which have in common their complexity, as well as an
inability to solve the problem. It is even difficult to agree what exactly is
the problem to be solved.''\\
\strut\hfill {\it (M. Lesieur: Turbulence in Fluids)}
\end{quote}

The above quote is the opening paragraph in one of the currently most widely
used monographs on fluid turbulence, \citeN{Lesieur}. It was not written having
solar granulation in mind ---but it could as well have been... Indeed, as we
will see in the next subsection, the most debated issues are the same in the
study of granulation as in other fields of turbulence: coherent structures and
intermittency ---those features that most strikingly distinguish turbulent flows
observed in the real world from those smooth, featureless random media that one
tends to envisage based on the phenomenological picture of the turbulent
cascade.

In his book Lesieur suggests 3 criteria for turbulence: (1) Randomness (2) An 
order-of-magnitude increase in macroscopic transport, in particular in the
transport of momentum\footnote{The macroscopic diffusion coefficients in
turbulence are $\sim LV$ (cf.\ Sect.~4).
The increase in the transport of momentum then implies the usual condition
$\Reyno\equiv LV/\nu\gg 1$} (3) The presence of motions on a wide and
continuous  range of spatial scales, spanning several decades.

The flows one tends to envisage on the basis of the above phenomenological 
picture of isotropic turbulence fulfil these expectations. But these
criteria admit a much wider class of random flows: in fact most of the
turbulent flows encountered in nature or in laboratories, while still
conforming to these definitive criteria, have a strikingly different appearance
from our na\"{\i}ve expectations. In particular, we note the ubiquitous
presence of {\it coherent structures\/} and {\it internal intermittency.}

Coherent structures are formations reminiscent of unstable normal modes  of the
instability feeding the turbulence. A classic example are coherent
Kelvin-Helmholtz-like vortices in the turbulent mixing layer. What is surprising
is their presence and longevity even in the regime of fully developed
turbulence. Individual coherent structures are not predictable, their place and
size varies with time and among different realizations of the flow. The presence
and characteristics of coherent structures cannot be foreseen on the basis of
spectral turbulence theory as these structures arise owing to the non-randomness
of relative phases of individual Fourier modes in the flow while power spectra
such as $E_k$ contain no phase information. 

The term ``internal intermittency'' refers to the fact that the dissipation
rate is not uniformly distributed over the flow volume; instead, the
dissipation is concentrated in an intermittent subset of the volume. This is
sometimes somewhat vaguely formulated saying that the flow ``is not everywhere
turbulent'', meaning that the energy cascade reaching down to the smallest
(dissipative) scales is only present in a (fractal) subset of space. Internal
intermittency also modifies structure functions and spectral exponents, in
particular for higher order correlations.

\subsection{Solar observations vs.\ theory}

The overall properties of the solar photosphere were recently reviewed on
these pages by \citeN{Solanki:SSR}. In what follows we will focus on
velocity fields only.  Spatio-temporal power spectra of velocity (or intensity)
fluctuations over the solar disk reveal that photospheric motions can be
divided into two broad classes. (See e.g. Fig.~1 in \citeNP{Straus+}.) On the
high-frequency side of the line corresponding the Brunt-V\"ais\"al\"a frequency
in the $k$--$\omega$ diagram one finds the familiar ``ridges'' corresponding to
the mostly discrete spectrum of solar oscillations. On the low-frequency side
of that line lies a broad continuum peaking at frequencies below 1 mHz and
wavenumbers below 2 Mm$^{-1}$. This continuum, corresponding to granular, meso-
and supergranular motions is what, guided by Lesieur's criterion (3) above, we
may tentatively identify with photospheric turbulence. 

The strongest intensity contrast and the highest velocity amplitude corresponds
to the characteristic cellular pattern of {\it granulation,} with typical
spatial and temporal scales of 1 Mm and 1000 s and a velocity amplitude of 1--2
km/s. The larger scale {\it mesogranulation\/} (10 Mm, $10^4$ s, 0.5--1 km/s)
and {\it supergranulation\/} (30 Mm, $10^5$ s, 0.3--0.5 km/s) are most easily
seen in the horizontal flow pattern or in the distributions of certain tracers
(magnetic flux tubes, exploding granules etc.). This suggests (despite
occasional views to the contrary) that {\it granulation represents the energy
input range of  photospheric turbulence\/} as its strong temperature
fluctuations inevitably imply strong buoyant acceleration. 

\subsubsection{Is granulation turbulence?}

It must be clear from the outset that these photospheric motions must be very
different from the textbook case of incompressible isotropic hydrodynamic
turbulence. In particular, the scale heights of pressure and density in the
photosphere are of the order of 100 km, i.e. ten times smaller than the supposed
energy input range! Besides, our observations cannot resolve features below
about 200 km on the solar surface; and granular observations are significantly
contaminated by seeing effects already well above the formal resolution limit
(\citeNP{Collados+Vazquez}). As this resolution limit is less than a decade
below the energy input scale one cannot reasonably hope to catch even a glimpse
of the presumed inertial range. And yet, many observers (e.g.\
\citeNP{Espagnet+}; \citeNP{Ruzmaikin+:supdif}) try, and often claim, to fit
granular power spectra by power laws or even Kolmogorov laws (ignoring not only
the seeing effects and the extreme inhomogeneity but also the complete
uncertainty regarding the theoretical form of the energy spectrum in 3D MHD
turbulence, cf.\ Section 2.1.2 above).

While such efforts are clearly futile, their futility hardly warrants taking the
extreme standpoint of denying the turbulent nature of granulation, as suggested
recently by \citeN{Nordlund+:provo}. These latter authors rightly criticize the
na\"{\i}ve attempts to fit granular power spectra by power laws, pointing out
that the power spectrum of a simple step function is $\sim k^{-2}$, a power law
that is hard to distinguish from the Kolmogorov law with the observational
errors at hand. Cell-like granules, with their rather sharp boundaries, can also
be approximately represented by a step-like function with steps of alternating
sign, thereby giving rise to a power-law-like spectrum. They then nicely
demonstrate that the very existence of well-defined, quasiregular granules is
due to non-random relative phases of the Fourier modes of the flow. However,
they then go on to suggest that this non-randomness of the phases and the
corresponding semiregular, cellular pattern of the granules, their more laminar
internal flow, together with the concentration of small-scale turbulence and
(presumably) dissipation in the intergranular lanes are at odds with a turbulent
nature of the granular flow. 

Indeed ---is granulation turbulence? Applying the criteria suggested by Lesieur
(Sect.\ 2.1.4 above) we can aswer affirmatively to this question. Granular
motions are obviously unpredictable, and they lead to an order-of-magnitude 
increase in magnetic diffusivity (cf.\ Sect. 3 below). From the supergranular
scales down to the resolution limit, a wide continuum of scales of motion is
present. Small-scale turbulence, especially below the resolution limit, may
indeed be concentrated in the intergranular lanes, together with dissipation
(\citeNP{Nesis+:granul.shear}) ---but this is nothing else than the phenomenon
of internal intermittency, well known from other types of turbulent flows. And
as for the non-random relative phases of Fourier modes, leading to the
organized, cellular appearance of granulation: this is the exact analogue of
the presence of coherent structures in other flows. Indeed, the idea I would
like to promote here is that {\it granulae are but one example of the extensive
family of coherent structures in turbulent flows}.

Whether or not we wish to call a certain flow turbulent is up to some point a
merely semantic issue. But excluding a flow type from among turbulent flows just
because of the presence of coherent structures and internal intermittency would
amount to exclude practically all flows realized in nature, or indeed even in
laboratory. Such a definition would hardly be of any practical use. In
conclusion, then: granulation is a perfectly typical turbulent flow ---typical
even in its atypicality.

\subsubsection{The problem of the size of granules}

The characteristic granular size of $\sim 1000\,$km presents a riddle as there
is no simple explanation for the preference of such a length scale. The
correlation length in deep turbulent convection is known to be approximately
equal to the pressure scale height (\citeNP{Chan+Sofia:Science}) while in the case of
granulation they differ by one order of magnitude. In the model of 
\citeN{Antia+:supgr} the scale is set by the scale of the most unstable normal
mode in a linear stability analysis of a mixing-length model of the convective
zone, taking into account turbulent heat transfer. The resulting scale is indeed
in reasonable agreement with granular sizes.

An alternative explanation was proposed recently by \citeN{Rast:SacPeak}. In
his scenario the basic units are not the granules but the downflow plumes in
between, initiated by localized cooling events. In 2D numerical experiments in
an adiabatic layer he finds that if the separation of two neighbouring plumes
exceeds a certain limit, a new starting plume spontaneously forms in between.
The relation of this result to the existence of two granule populations
(\citeNP{Hirzberger+}) is unclear but the mechanism suggested by Rast, coupled
with the supposed coalescence of plumes getting too close, could regulate the
mean separation between plumes. It remains to be seen whether such a scenario
can also reproduce granule sizes quantitatively, in a realistic, 3D model.

\subsubsection{The origin of supergranulation}

If the energy deposited at the granular scales cascades towards the smaller
scales, to be ultimately dissipated in the intergranular lanes, what is the
origin of larger scale velocity fields such as meso- and supergranulation? 
Maybe the oldest answer to this question is that they may be the surface
imprint of larger scale convective motions taking place in deeper layers where
the pressure scale height is much larger. In their work cited above, 
\citeN{Antia+:supgr} found a second peak in the growth rate--wavenumber
relation of normal modes. The scale of this peak, 10--20 Mm, invites
identification with supergranulation. 

Another suggestion was made by \citeN{Krishan:inverse} who proposed that
supergranulation is due to some kind of inverse cascade taking place in the
photosphere. The particular cascade she proposed involved the mean square
kinetic helicity ---an ideal invariant in 3D hydrodynamic turbulence. 

Supergranulation is best seen in the distribution of network magnetic fields,
leading to the suggestion that the magnetic field itself may have a crucial
role in the origin of the flow pattern (Zwaan, private comm.). This scenario
may possibly be combined with an inverse-cascade-type model recalling that in
3D MHD turbulence magnetic helicity suffers an inverse cascade.

As we see, possibilities abound but none has been pursued to the point
of giving testable predictions. The problem of the origin of supergranulation is
as open as it was three decades ago, at the time of its discovery.

\section{The Turbulent Magnetic Field}

\subsection{$\alpha$-Effect and Small-Scale Dynamo}

Beside the fluctuating velocity field, magnetic fields also have a fluctuating
component in a turbulent plasma. While a strong large-scale field will
obviously contribute to small-scale fluctuations, a turbulent magnetic field of
significant strength is present in 3D MHD turbulence even in the absence of a
global field. This field is generated by small-scale dynamo action. 

Dynamo effects come in two varieties. {\it Large-scale\/} dynamos involve the
growth/maintenance of an organized, large-scale (as compared to the integral
scale of turbulence) mean magnetic field. As we already mentioned, large-scale
dynamo effect is intimately linked to the inverse cascade of magnetic helicity
in 3D MHD turbulence. The best known of these effects is the {\it
$\alpha$-effect,} consisting of the appearance of a mean electromotive force
proportional to the mean field strength: $\vec{\cal E}=\alpha \langle\vc B\rangle$. 
$\alpha$ is a pseudoscalar, so for this effect to appear, the 
flow should violate parity invariance.  

In {\it small-scale\/} dynamos, in contrast, no mean magnetic field is
generated; yet, the mean magnetic energy density $\langle B^2\rangle/2\mu$ and
the unsigned flux density $\langle|\vc B|\rangle$, grow exponentially
(kinematical case) or are maintained (hydromagnetic case). This effect can also
occur in parity invariant flows.

Consider the most widely known (though by no means the only) physical mechanism
capable of maintaining a dynamo: helical flows. A localized helical flow acts a
stretch-twist-fold type dynamo on a preexisting weak large-scale field. The
resulting field configuration can be regarded as the superposition of a closed
flux ring and the original weak field (on which the whole process can be
infinitely repeated). Whether the closed loop is actually detached from the
preexisting field by reconnection or this is just a convenient mathematical
decomposition, is irrelevant. The orientation of the ring depends on
the sign of the helicity in the original flow. If the mean helicity is
zero (mirror symmetry) the resulting loops will be
randomly oriented and their field will cancel on the mean, i.e.\ no mean
magnetic field is produced. Their magnetic energy density however does not
cancel so we are dealing with a small-scale dynamo. If, on the other hand,
there is a preferred sense of rotation in the flow (mean helicity does not
vanish) the loops will show a preferred orientation thus giving rise to a net
large-scale field: this is the case of the large-scale dynamo. 

The condition for these effects to work in a flow is in general a sufficiently
high magnetic Reynolds number $LV/\nu_m$ where $\nu_m$ is the (molecular) magnetic
diffusivity. This condition is fulfilled in the solar photosphere, so it has
been suggested by several authors (\citeNP{Petrovay+Szakaly:AA1};
\citeNP{Cattaneo:small.dyn}) that a small-scale dynamo is operative there.
(Mean helicity is negligible in the photosphere so large-scale dynamo effect is
absent.)

What is the saturated state of the small-scale dynamo? Numerical experiments 
and turbulence closure models (e.g.\ \citeNP{DeYoung};
\citeNP{Meneguzzi+Pouquet:JFM}; \citeNP{Cattaneo:small.dyn};
\citeNP{Subramanian:unified}) invariably  show that in high Reynolds number
driven non-helical MHD turbulence, especially in flows of convective origin,
\begin{lista}
\item magnetic energy density saturates at a value about an order of magnitude
below kinetic energy density
\item magnetic field structure is extremely intermittent; the strongest 
structures (tubes) have a magnetic energy density comparable to the kinetic
energy density; the typical tube filling factor is thus $f\sim 0.1$. The
analytic result of \citeN{Subramanian:unified} predicts $f=0.025\,TV/L$,
$T$ being the correlation time.
\item the magnetic energy spectrum peaks at significantly higher wavenumbers 
than the kinetic energy spectrum. 
\end{lista}

An issue that has given rise to much  controversy in recent years has been the
claimed strong suppression of the $\alpha$-effect by even a very weak magnetic
field (see e.g.\ Section~4 of \citeNP{Petrovay+Zsargo}, for a summary of the
argument). Here I would just like to call attention to the little appreciated
fact that if a suppression of $\alpha$ does take place then it should also
affect the small-scale dynamo, given that the underlying physical mechanisms
are the same. While the large-scale dynamo may be salvaged by other
non-classical $\alpha$-mechanisms, this may be much more difficult for the
small-scale dynamo. And yet, as we said, small-scale dynamo action undeniably
exists and is extremely common in turbulent MHD simulations. This  puzzle is
not just unsolved but rarely even recognized. 

On the basis of what was said above one may expect that the photospheric
turbulent field should be characterized by a mean unsigned flux density that is
crudely an order of magnitude lower than the photospheric equipartition field
strength (400 G), and its characteristic scale (correlation length) should be
about 100\,km, well below the resolution limit. How can we hope to observe
this mixed-polarity field?

\subsection{Hanle Diagnostics of the Turbulent Magnetic Field}

The Zeeman effect, our traditional diagnostic tool for solar magnetism, is
``blind'' to fields with polarities mixed on scales below the resolution limit.
In such a field the net magnetic flux in a pixel most of the time remains below
the resolution limit, and only occasionally, thanks to statistical
fluctuations, can one expect to detect a weak signal of random polarity. This
Zeeman-detectable part of the turbulent field, the so-called {\it intranetwork
field\/} is however but the tip of the iceberg, and most of the turbulent field
remains undetected.

An alternative detection method is offered by the {\it Hanle effect.} This
effect consists in a reduction of the linear polarization of coherently
scattered light if the scattering occurs in a magnetic field
(\citeNP{Stenflo:book}). The use of this method for the detection of solar
turbulent fields has come to its own in the last decade (\citeNP{Faurob};
\citeNP{Bianda+:Sr}, etc.). The increased interest in a rigorous interpretation
of scattering line polarization observations has motivated the development of
very efficient numerical methods for the transfer of polarized radiation and
Hanle-effect codes (\citeNP{JTB+RMS}). The consensus now
seems to be that at the height of line formation the turbulent flux density is
$\sim$ 10-20 G; however, many controversies still remain, such as the ``Na
solar paradox'' (\citeNP{Landi:Nature}). For a critical discussion on this
point, with an advance of some interesting new results see \citeN{JTB:2nd.sp}.


One of the most intriguing recent discoveries is the finding of
\citeN{Stenflo+:Hanle.fluct} that the degree of linear polarization $Q/I$ shows
large amplitude random variations over the solar disk. A phenomenological model
for the origin of such fluctuations (\citeNP{Petrovay:tfluc}) suggests that they
may be interpreted rather naturally on the basis of statistical fluctuations of
the small-scale dynamo mechanism.

\section{Turbulent magnetic diffusion}


\subsection{Random walk and Babcock-Leighton models}

The random horizontal flow on the solar surface leads to a random motion of
tracers, the most important of which are magnetic flux tubes. A simple random
walk of stepsize $\Delta x$ and timestep $\Delta t$ over a plane is known to
lead to an increase of the rms separation $r$ of a tracer from its starting
point (or of two tracers from each other) according to the law
\begin{equation}
 r^2=4Dt \label{eq:Fick} 
\end{equation}
where $D=\Delta x^2/2\Delta t$. The time development of a continuous 
distribution of tracers is then described by a diffusion equation with
diffusivity $D$.

As a first approximation, the advection of tracers by (super)granules
may be represented by such a simple random walk/diffusion, identifying $\Delta
x$ with the spatial scale $l$ of the cells and $\Delta t$ with their lifetime
$\tau$. 


This is the approach used in Babcock--Leighton-type models of the solar cycle
where poloidal fields are brought to the surface in a concentrated form in
active regions, and thereafter they are passively transported to the poles by
transport processes (diffusion and meridional circulation). The diffusivity in
these models is a free parameter: a best fit to the observations yields
$D=600\,$km$^2/$s.  Despite the vectorial character of the magnetic field,
these 1D models have been remarkably successful in reproducing the observed
temporal evolution of the flux distribution. A possible explanation was
proposed in their model by \citeN{Wang+:1.5D}: they assume that field lines are
vertically oriented throughout much of the convective zone and this essentially
reduces the problem to one dimension. Some support for this conjecture has
come from the 2D flux transport models of \citeN{Petrovay+Szakaly:SPh}. Thus,
in a first approximation, 1D models may be used for the description of
meridional transport, as these fields pervade the convective zone and are
continuously reprocessed through it.

The empirically determined value of the diffusivity, 600\,km$^2$/s, seems to
agree with the primitive random walk model if the steps are identified with
granular sizes/lifetimes. Supergranulation, however, should lead to a
diffusivity that is by an order of magnitude higher than this calibration. The
continuous reprocessing of large-scale fields throughout the convective zone
offers a plausible explanation for this inconsistency: the empirical value of
the diffusivity reflects the turbulent diffusivity in the lower convective zone
where the pressure scale height is $\sim$50 Mm and the turnover time $\sim$1
month.

An alternative explanation was put forward by \citeN{Ruzmaikin+:cellular} who
pointed out that, owing to the cellular nature of photospheric flows,
identifying cell size $l$ and lifetime $\tau$ with random walk steps is an
oversimplification. The fact that a tracer cannot leave a cell during the cell's
lifetime, even if it was originally placed next to its border, reduces the
effective stepsize significantly. The resulting reduction in $D$ is very
sensitive to the value of the Strouhal number $\Strouno=\tau v/l$ and is
especially strong for $\Strouno\gg 1$. This effect
may be sufficient to reduce supergranular diffusivity to the observed value.

\subsection{Anomalous diffusion} 

The cellular and turbulent nature of the flow implies that a simple random walk
cannot account for the motion of magnetic elements. As a consequence, the
actual flux redistribution may differ from a simple (Fickian)  diffusion
process (\citeNP{Isichenko}) and, instead of equation (\ref{eq:Fick}) in
general one has 
\begin{equation}
 r= 2 K t^\zeta .  \label{eq:nonFick}    
\end{equation} 
If $\zeta\neq 1/2$ it is customary to speak of {\it anomalous\/} or {\it
non-Fickian diffusion,} $\zeta>1/2$ corresponding to {\it superdiffusion,}
$\zeta<1/2$ to {\it subdiffusion.} As equation (\ref{eq:nonFick}) means
a unique relation between $r$ and $t$ one might formally still write $r=2 K'(r)
t^{1/2}$, leading to the concept of a ``scale-dependent diffusivity''
\begin{equation}
 D(r)=K'^2=K^{1/\zeta}r^{2-1/\zeta} \label{eq:scaledepdif}    
\end{equation}  It is
however clear that such a concept is in general useless for the description of
the  evolution of a continuous field where no preferred scale exists. Anomalous
diffusion thus cannot be described by a diffusion equation or, indeed, by any
partial differential equation. 

How can anomalous diffusion come about? One possibility was suggested by
\citeN{Schrijver+Martin:}. Magnetic flux tubes are
located at junctions of a fractal lattice between supergranules,
mesogranules and granules. Assuming that limitations exist for the motion
of individual flux elements along this lattice, for certain lattice properties
subdiffusion may result. They made an attempt to detect subdiffusion by the
analysis of observed flux redistribution in the photosphere; however, $\zeta$
was not found to differ from 0.5 within the observational uncertanties.

Being a multiscale phenomenon, turbulence can also naturally lead to a
``scale-dependent diffusivity''. In order to understand the nature of the
diffusion process in a turbulent medium let us consider the question how a
random continuous velocity field of a given characteristic scale $\lambda$
(i.e. one level in the turbulent hierarchy) can be best represented by a random
walk with steps $\Delta x$ and $\Delta t$. For the best representation one
should set $\Delta t=\tau_L$, the Lagrangian correlation time of the flow, as
this is just the time after which the advected particle experiences a
significant change in its velocity. The distance the particle travels in this
time is $\Delta x=v\tau_L=\min(v\tau_E, \lambda)$ where $\tau_E$ is the
Eulerian correlation time, $\lambda$ the correlation length, and $v$ the rms
velocity. The diffusivity for this random walk will thus depend on the Strouhal
number $\Strouno=\tau_E v/\lambda$; assuming a non-cellular flow  
\begin{equation}
 D=\left\{\begin{array}{ll}\tau_E v^2 &\qquad \mbox{if } \Strouno <1\\
                      \lambda v  &\qquad \mbox{if } \Strouno > 1
     \end{array}\right.
     \label{eq:Dexpr}  
\end{equation}

In a multiscale flow both $\tau_E$ and $v$ scale with $\lambda$:
\begin{equation}
 \tau_E\sim \lambda^z \qquad v^2\sim\lambda^{\alpha-1} \label{eq:scalings} 
\end{equation}
During the random walk, motions on scales exceeding the separation $r$ of two
tracers do not contribute to their further separation while all other scales
contribute to it. Of these scales, according to equation
(\ref{eq:scaledepdif}), the smallest one will dominate in the diffusion process
for $2-1/\zeta<1$ i.e.\ $\zeta<1/2$. In this case, then, the diffusivity will
not significantly depend on the separation for all scales above the viscous
scale: turbulence can never lead to subdiffusion.

\begin{figure}
\centerline{\psfig{figure=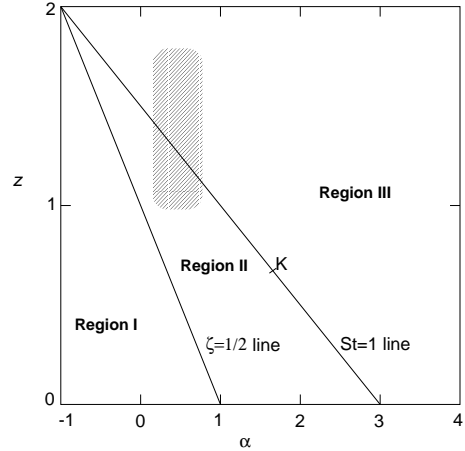,width=6 cm}}
\caption{Regimes of anomalous diffusion on the $\alpha$--$z$ plane. $K$ is the
Kolmogorov point; the shaded area indicates the approximate position of
the photospheric flow field in the 1--30 Mm size range. (Non-cellular case)}
\label{fig:zalpha}
\end{figure}

In the case when the relatively largest scale $\lambda\sim r$ dominates,  for
given values of $z$ and $\alpha$ the anomalous diffusion exponent $\zeta$ can
be determined by substituting (\ref{eq:scalings}) into (\ref{eq:Dexpr}) and
equating the resulting scaling exponent of $D$ to $2-1/\zeta$, as given by
(\ref{eq:scaledepdif}). The Strouhal number scales as
$\Strouno\sim\lambda^{z+\alpha/2-3/2}$. For high Reynolds numbers, then, the
sign of $(\Strouno-1)$ at the larger scales depends on the sign of the scaling
exponent of $\Strouno$, i.e.\ the line
\begin{equation}
2z+\alpha-3=0 
\end{equation}
defines two regimes in the $\alpha$--$z$ plane (Fig.~\ref{fig:zalpha}). Above
the line, in what is called Region III (\citeNP{Avellaneda+Majda:review}), one 
finds
$\zeta=2/(3-\alpha)$ (except in the case of a cellular flow when $\zeta=1/z$
---cf.\ the discussion at the end of Sect.\ 4.1). This Region is clearly 
superdiffusive for all values of $\alpha>-1$ (or $z<2$). Below the line, in
Region II we have $\zeta=1/(3-z-\alpha)$, independent of cellularity, as 
here we have low Strouhal numbers at the large scales. It is then clear
that a second dividing line will also exist at
\begin{equation}
z+\alpha=1   
\end{equation}
as below that line (Region I) $\zeta<1/2$ would result, in which case, as we
have seen, the smallest scales dominate the diffusion process. Region I is thus
characterized by a Fickian diffusion, while Region II is again superdiffusive.
Point $K$ in Figure 1 denotes the case of a Kolmogorov spectrum, $\alpha=5/3$,
$z=2/3$, $\zeta=3/2$.

In order to determine the place of photospheric velocity fields on the
$\alpha$--$z$ diagram, \citeN{Ruzmaikin+:supdif} fitted power laws to the 
spatio-temporal power spectra of photospheric velocity fields with the result 
$\alpha\sim 1.5$--$1.8$, $z\sim 0.15$--$0.85$. This would localize solar
turbulence to the neighbourhood of the Kolmogorov point $K$. However, in Section 2.2.1 above we already stressed the
perils of power-law fits to power spectra of solar velocity fields. There is
simply no theoretical reason or observational evidence to suggest that these
fields should follow a power-law spectrum from supergranular scales down to the
resolution limit. Indeed, the well known fact that meso- and supergranular
motions have a lower velocity amplitude than granulation, tells us that
$\alpha<1$ in the regime $\lambda>1\,$Mm! Using observational estimates for
these velocity amplitudes and for correlation times one arrives at much more
robust limits that are in plain contradiction to the ones quoted above:
$\alpha\sim 0.0$--$0.7$, $z\sim 0.9$--$1.8$, leading to $\zeta\sim 0.48$--$1.2$.
These limits in themselves would indicate superdiffusion (shaded area in 
Fig.~\ref{fig:zalpha}).

Turbulent erosion models of sunspot decay   can also be used to constrain
anomalous diffusion in the photosphere (\citeNP{Petrovay:supdif}). The size of
sunspots spans the granular-supergranular size range that is of interest in
this respect, and the fortuitous property of the erosion models that they {\it
do\/} show a characteristic scale, the radius of the spontaneously formed
current sheet, makes it possible to test for $\zeta$ by using a scale-dependent
diffusion coefficient with the current sheet radius as defining scale. In this
way, $\zeta$ is found to lie in the range $0.44$--$0.59$, i.e.\ any deviation
from a Fickian diffusion seems to be modest, if present at all. A possible
explanation for why the diffusion exponent is lower than suggested by velocity
power spectra may be that superdiffusion due to turbulence is offset by
subdiffusion effects due to diffusion on a bond lattice.

\acknowledgements 
This work was funded by the DGES project no.~95-0028, the FKFP project 
no.~0201/97, and the OTKA project no.~T032462.



\end{article}
\end{document}